\documentclass{article}
\usepackage{iclr2025_conference,times}

\usepackage{amsmath,amsfonts,bm}

\def\eqref#1{equation~\ref{#1}}

\def\1{\bm{1}}

\DeclareMathAlphabet{\mathsfit}{\encodingdefault}{\sfdefault}{m}{sl}
\SetMathAlphabet{\mathsfit}{bold}{\encodingdefault}{\sfdefault}{bx}{n}

\usepackage{hyperref}
\usepackage{url}
\usepackage{graphicx}
\usepackage{booktabs}
\usepackage{multirow}
\usepackage{amsmath}
\usepackage{xcolor}

\title{Towards Large-Scale Heterogeneous Data Organization for Scientific Foundation Models: A Nuclear Fusion Case Study}

\author{Nathaniel Chen\\
Mechanical and Aerospace Engineering\\
Princeton University\\
Princeton, NJ 08544, USA \\
\texttt{nathaniel@princeton.edu} \\
\And
Kouroche Bouchiat\\
Mechanical and Aerospace Engineering\\
Princeton University\\
Princeton, NJ 08544, USA \\
\And
Peter Steiner \\
Princeton University\\
Princeton, NJ 08544, USA \\
\And
Azaraksh Jalalvand \\
Princeton University\\
Princeton, NJ 08544, USA \\
\And
Sangkyeun Kim \\
Princeton University\\
Princeton, NJ 08544, USA \\
\And
Egemen Kolemen \\
Princeton Plasma Physics Laboratory\\
Princeton University\\
Princeton, NJ 08544, USA \\
}

\iclrfinalcopy 
\begin{document}

\maketitle

\begin{abstract}
Training effective foundation models requires massive and organized datasets, yet scientific domains such as nuclear fusion present unique challenges due to largely heterogeneous and sparse data. Here we characterize the data used in developing such a model: with over 20 sensor types spanning 5 orders of magnitude in sampling rate, mixed tensor structures (point measurements, spectrograms, images), and nonstationary physics. We analyze our input complexity and discuss trade-offs between temporal context and frequency resolution. Our analysis provides a template for representing multi-modal fluctuation data at scale, with implications for both multi-modal control systems and nuclear fusion.
\end{abstract}

\section{Introduction}

Foundation models have transformed machine learning via their ability to learn general representations from large-scale data~\citep{bommasani_opportunities_2021}. While scaling laws are well-characterized for language and vision~\citep{kaplan_scaling_2020, zhai_scaling_2022}, scientific domains present distinct data organization challenges that remain.

Tokamak data offers an interesting case study. Experiments can generate rich multivariate text, time series, and images, measuring fusion behavior across spatial, temporal, and spectral dimensions~\citep{churchill_ai_2025}. The data is inherently heterogeneous as sensors can sample from a range of low frequency amplitude-focused scalars to high frequency fluctuation measurements~\citep{gohil_charge_1991}. This data can also be sparse, in which enabling certain actuators during experiments can disable nearby sensor pickup even if data is still present. Heterogeneity, combined with turbulent non-stationary physics, makes standard data loading difficult~\citep{surko_waves_1983}.

Prior work has addressed fusion prediction tasks individually: disruption forecasting~\citep{vega_disruption_2022}, profile prediction~\citep{char_full_2024}, and cross-diagnostic inference~\citep{abbate_large-database_2024, jalalvand_multimodal_2025}. More recent studies have investigated methods of reducing the latent space of spectral data~\citep{bouchiat_towards_2025}. However, each approach employs task-specific preprocessing, hindering the development of a scalable dataset for standardized models.

Here, we characterize fusion data heterogeneity and its implications for foundation model training. First, we categorize 23 candidate measurements from the DIII-D tokamak by sampling rate, tensor structure, and role in physics; second, we show approximate scaling relationships relating window size to model input complexity; third, we identify key design trade-offs for data loading pipelines.

\section{The Fusion Data Landscape}
\subsection{Shared Measurements and Input Estimations}
We analyze sensors and actuators from across several tokamaks, namely DIII-D and guided by similar structures in other devices including KSTAR which are representative of modern fusion facilities~\citep{fenstermacher_diii-d_2022, lee_design_1999}. A proposed method of enabling cross-device modeling with the same architecture can use relative encoding for sensor arrays, which are shared characteristics across devices. For example, interferometers on both DIII-D and KSTAR form a common multichannel configuration~\citep{carlstrom_realtime_1988,lee_design_2016}, while electron cyclotron emission (ECE) often follows a linear array structure~\citep{austin_electron_2003, jeong_electron_2010}. Magnetics also often follow a gridded pattern, though in toroidal coordinates instead of Cartesian~\citep{strait_magnetic_2006}. While device sizes and operating magnetic field strengths can greatly influence the range of observed phenomena, measurement rates across devices are often similar.

Table~\ref{tab:data_summary} summarizes the heterogeneous structure.

\begin{table}[ht]
\centering
\caption{Heterogeneous fusion data: 23 diagnostics across 6 physics categories which fall into 12 more general categories given a 100ms window.}
\label{tab:data_summary}
\vspace{0.5em}
\resizebox{\columnwidth}{!}{%
\begin{tabular}{llrrr}
\toprule
\textbf{Category} & \textbf{Representative} & \textbf{Rate (Hz)} & \textbf{Shape} & \textbf{Elements} \\
\midrule
Fluctuations & Magnetics, BES & 100--2000 & $(C, F, T)$ & 365,714 \\
Kinetic & Thomson, CER & 0.04--0.2 & $(C, T)$ & 6,089 \\
Equilibrium & MSE, $I_p$ & 1.5--2.5 & $(C, T)$ & 3,133 \\
Heat/Fuel & ECH, NBI, Gas & 2--100 & $(C, T)$ & 2,605 \\
Radiation & ECE, $D_\alpha$ & 50--1667 & $(C, F, T)$ & 136,867 \\
Imaging & Divertor Camera & 1 & $(H, W, 1)$ & 65,536 \\
\midrule
\multicolumn{4}{r}{\textbf{Total (100ms window)}} & $\mathbf{\sim}$\textbf{398,500} \\
\bottomrule
\end{tabular}%
}
\vspace{-1em}
\end{table}

Some key observations include:

\begin{enumerate}
    \item \textbf{Extreme sampling rate diversity.} Rates span roughly 5 orders of magnitude (0.04 Hz for Thomson scattering to 2000 Hz for high-resolution magnetics)~\citep{ponce-marquez_thomson_2010,strait_magnetic_2006}. This discourages naive concatenation between measurements since resampling to a common rate either discards high-frequency physics or explodes low-frequency channel dimensions.
    \item \textbf{Mixed tensor structures.} Data includes 2D time series $(C, T)$, 3D spectral tensors $(C, F, T)$, and images $(H, W, 1)$~\citep{chen_regulation_2026}. Unlike vision (fixed $224 \times 224 \times 3$) or language (variable-length 1D sequences), fusion models may require architectures to handle heterogeneous tensor ranks simultaneously which is more akin to time series foundation models~\citep{talukder_totem_2025}, but with a required accuracy of audio models~\citep{kumar_high-fidelity_2023}. This encourages the adoption targeting spectral or video data.
    \item \textbf{Spectral dominance.} Magnetic and density fluctuation spectrograms comprise 87\% of total elements, yet represent only 5 of 23 unique modalities. This imbalance may affect both compute allocation and representation learning since models may overfit to spectral features while underweighting scalar quantities critical for understanding physics.
    \item \textbf{Noise and Non-stationarity.} Thermal noise can greatly obscure coherent structures. Turbulent transport causes statistical properties to evolve on $\sim$10ms timescales, while global equilibrium evolves over $\sim$1s~\citep{luce_development_2005}. Standard normalization (z-score, min-max) computed over full discharges may fail to capture this multi-scale structure.
\end{enumerate}

\subsection{Stored }
\begin{figure}
\centering
\includegraphics[width=\columnwidth]{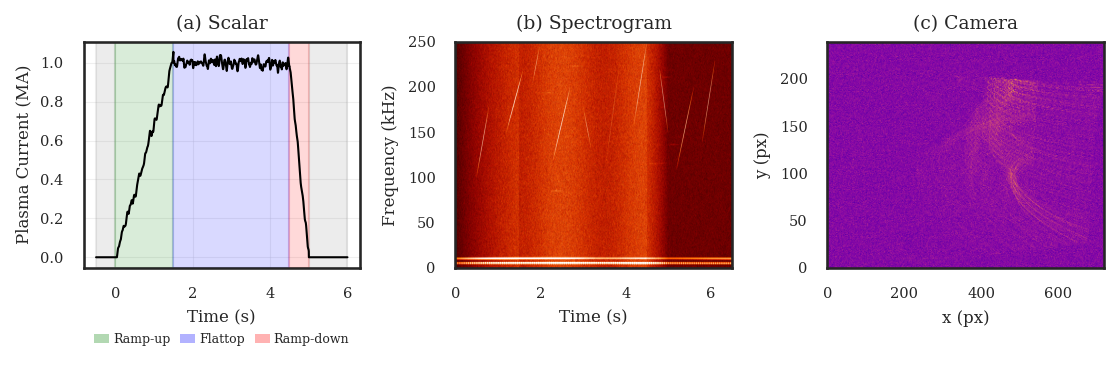}
\caption{Synthetic examples of the three primary data modalities in fusion diagnostics: (a) representative scalar time series (plasma current), (b) spectrogram (frequency vs.\ time) where chirps and low frequency modes can both be present at a given time, and (c) camera images in which pixel-level changes can indicate different physics.}
\label{fig:synthetic}
\vspace{-1em}
\end{figure}

Measurement identification important to prediction are often sparse and difficult to observe. For illustrative clarity, synthetic data is shown to represent the kind of data often analyzed in the context of nuclear fusion (Figure~\ref{fig:synthetic}).

Current treatment of these signals broadly categorize them into 3 types of representations.

Scalar representations are used in areas where amplitude information is more important. These can include measurements geared towards 

Spectral representations are used in areas where frequency information can be more important. While many methods exist, the Short Time Fourier Transform (STFT) is used due to its versatility.

Camera representations.

In addition, there are actuator signals.

\section{Design Considerations}

\subsection{Window Size Trade-offs}

The choice of our time window $w$ fundamentally shapes the learning problem. The frequency resolution trade-off may play an important role in the captured dynamics:
\begin{itemize}
    \item \textbf{Alfv\'{e}n eigenmodes} (100-250 kHz): Requires $w \geq 1$ms for adequate spectral resolution, but benefit from longer context to track frequency chirping~\citep{zeeland_alfven_2005}.
    \item \textbf{Tearing modes} (1--20 kHz): Evolve slowly, permitting $w \geq 100$ms windows that capture mode locking dynamics~\citep{la_haye_control_2002}.
\end{itemize}

\begin{figure}
\centering
\includegraphics[width=\columnwidth]{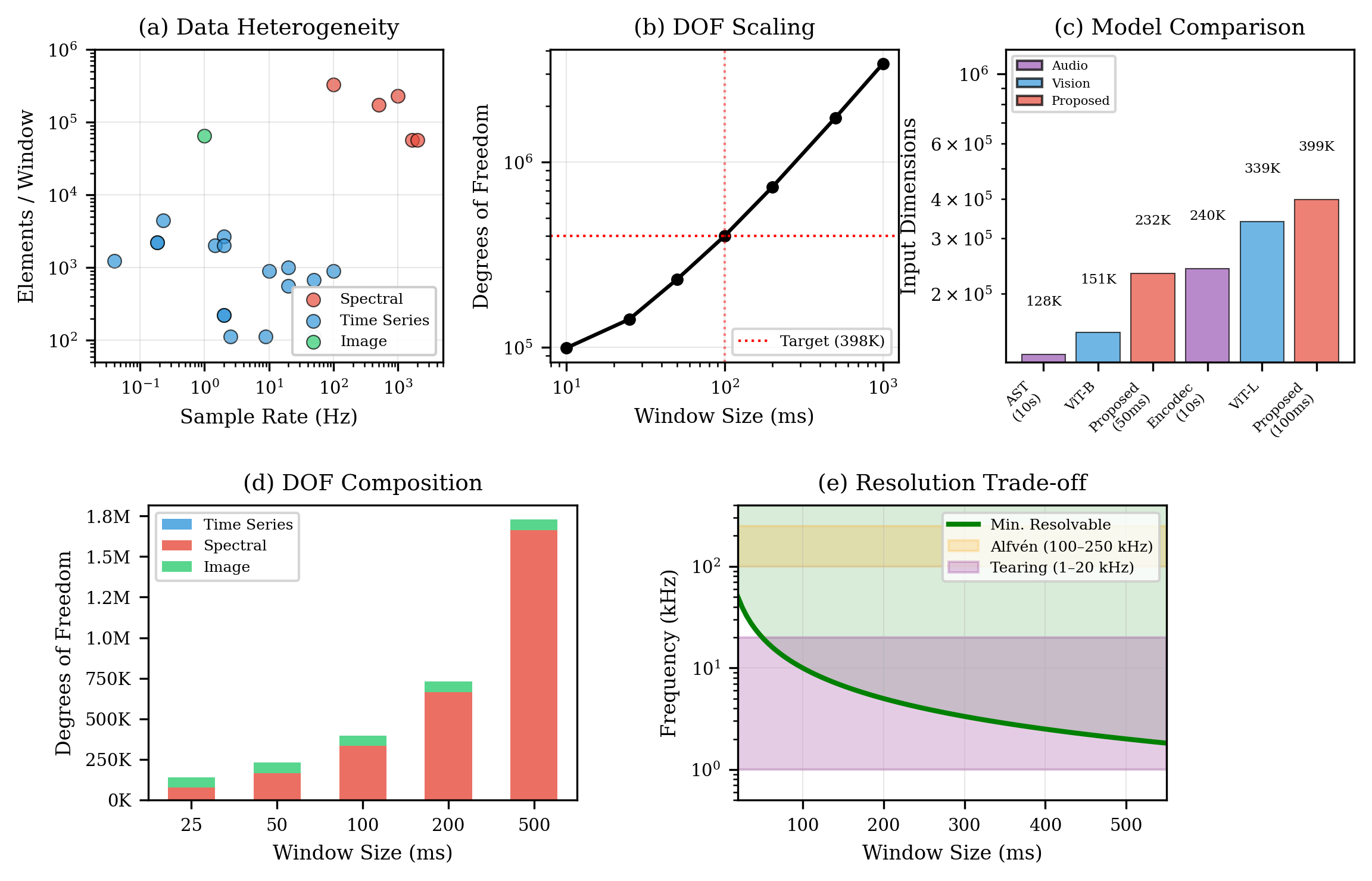}
\caption{Window size analysis. (a) Data heterogeneity across sampling rates and modalities. (b) Degrees of freedom scale approximately linearly with window size. (c) Model input dimension comparison with existing audio and vision architectures~\citep{gong_ast_2021, baevski_wav2vec_2020,dosovitskiy_image_2020}. (d) DOF composition by data type. (e) Frequency resolution trade-off: larger windows capture lower frequencies but sacrifice temporal localization.}
\label{fig:scaling}
\vspace{-1em}
\end{figure}

Based on task requirements in prior work, a $50$ms baseline is recommended for modeling a good amount of the kinetics. However, given a major goal of autoregressive prediction, this would require 100 accurate prediction steps for an average shot that is $5000$ms. Increasing the window to $w = 100$ms as a baseline halves the auto-regressive roll-out depth, potentially stabilizing training, yielding an input sequence length in the range of a large ViT model at $\sim$398K inputs.

\subsection{Precomputation vs.\ On-the-Fly Processing}

Given the current TB scale of combined fusion data on hand, we must decide what to precompute. An important consideration should be made for fluctuation measurements, especially depending on the architecture they are to be potentially used in (AST which requires patch spectrogram inputs vs.\ wav2vec which operates on raw waveforms)~\citep{gong_ast_2021,baevski_wav2vec_2020}.

\textbf{Precompute:} (1) Spectrograms from raw waveforms (reduces 2000 Hz $\rightarrow$ 500 Hz $\times$ 128 freq bins) with additional preprocessing such as clipping and filtering; (2) Spatial interpolation to common grids.

\textbf{On-the-fly:} (1) Window extraction and padding; (2) Normalization (requires batch statistics); (3) Augmentations (time shifts, channel dropout).

In addition, methods such as wavelet transforms and multiscale transforms have been shown to improve performance. However, these may introduce additional overhead for baseline tasks.

\subsection{Handling Sparsity and Missing Data}

Fusion diagnostics exhibit structured sparsity: sensors fail, diagnostics operate intermittently, and device configurations vary across facilities. We propose:
\begin{itemize}
    \item \textbf{Channel masking}: Positional embeddings can indicate the input and measurement ID, with a sparse input and reconstructed output with masked variables indicating that the measurement is not available, while keeping computation tractable~\citep{baade_mae-ast_2022}.
    \item \textbf{Cross-device alignment}: Mapping diagnostics to common categories (Table~\ref{tab:data_summary}) rather than device-specific names, facilitating transfer between DIII-D, KSTAR, EAST, and future ITER data. This also includes mapping the condition of the device during experiments, as documented in log files, which can greatly affect the results of experiments even with the same actuation and measurements.
\end{itemize}

\section{Conclusion}

We have characterized the heterogeneous data landscape of nuclear fusion and identified key challenges for foundation model training: extreme sampling rate diversity, mixed tensor structures, spectral data dominance, and physics-driven non-stationarity. Our window size scaling analysis and design recommendations provide a template for organizing scientific data at scale.

\textbf{Future work} includes training a foundation model to identify critical considerations: (1) developing full scaling laws across model size and data volume; (2) benchmarks on downstream tasks such as Alfv\'{e}n eigenmode classification, ELM prediction, and tearing mode forecasting; (3) cross-device transfer experiments.


\bibliography{iclr2025_conference}
\bibliographystyle{iclr2025_conference}

\end{document}